\documentclass[aps,floats,superscriptaddress]{revtex4}
\usepackage{amsmath}

\usepackage{epsfig}
\usepackage{graphicx}
\usepackage{dcolumn}
\usepackage{bm}

\renewcommand{\dag}{^{\dagger}}
\newcommand{\tc}{\tau_{\rm c}}
\newcommand{\Dren}{\Delta_{\rm ren}}
\newcommand{\De}{\Delta}

\def\gapp{\lower.35em\hbox{$\stackrel{\textstyle>}{\sim}$}}
\def\lapp{\lower.35em\hbox{$\stackrel{\textstyle<}{\sim}$}}

\begin{document}
%

\title{
Entanglement of spin chains with general boundaries and of dissipative systems}
\author{T. Stauber} 
\affiliation{Centro de F\'{\i}sica  e  Departamento de
F\'{\i}sica, Universidade do Minho, P-4710-057, Braga, Portugal}
\author{F. Guinea}
\affiliation{Instituto de Ciencia de Materiales de Madrid, CSIC, Cantoblanco, E-28049 Madrid, Spain.}
\date{\today}
\begin{abstract}
We analyze the entanglement properties of spins (qubits) close to the boundary of spin chains in the vicinity of a quantum critical point and show that the concurrence at the boundary is significantly different from the one of bulk spins. We also discuss the von Neumann entropy of dissipative environments in the vicinity of a (boundary) critical point, such as two Ising-coupled Kondo-impurities or the dissipative two-level system. Our results indicate that the entanglement (concurrence and/or von Neumann entropy) changes abruptly at the point where coherent quantum oscillations cease to exist. The phase transition modifies significantly less the entanglement if no symmetry breaking field is applied and we argue that this might be a general property of the entanglement of dissipative systems. We finally analyze the entanglement of an harmonic chain between the two ends as function of the system size.
\end{abstract}
%
\pacs{03.65.Ud, 03.67.Hk}

%
%
%
%
\maketitle 
\section{Introduction}
Coherence, decoherence and the measurement process are longstanding
problems of quantum mechanics since they mark the fundamental
difference to classical systems. They have gained increasing importance
in the context of quantum computing because the operation of a quantum
computer requires a careful control of the interaction between the
system and its macroscopic environment. The resulting entanglement
between the system's degrees of freedom and the reservoir has been a
recurrent topic since the formulation of quantum mechanics, as it is
relevant to the analysis of the measurement process
\cite{O92,Z03,AmicoReview}. 

Theoretical research on Macroscopic Quantum Tunneling lead, among
other results, to the (re)formulation of a canonical model for the
analysis of a quantum system interacting with a macroscopic
environment, the so called Caldeira-Leggett model \cite{Cal83},
initially introduced by Feynman and Vernon\cite{FeyVer63}. It can be
shown that this canonical model describes correctly the low energy
features of a system which, in the classical limit, undergoes Ohmic
dissipation (linear friction). It can be extended to systems with more
complicated, non-linear, dissipative properties, usually called
sub-Ohmic and super-Ohmic, see below \cite{Letal87,W99}.

In relation to the ongoing research on entanglement, a recent
interesting development is the analysis of the concurrence of
spin-chains like the transverse Ising model and the XY model. It was found that the derivative of the concurrence obeys universal scaling relations close to the quantum critical point and eventually diverges at the transition 
\cite{Oetal02,ON02}. Also other models which
exhibit a quantum phase transition were subsequently investigated in this
direction, as e.g. the Lipkin-Meshkov-Glick model
\cite{Vidal04,DV05}. 

Originally, the concurrence as measure of entanglement was
introduced by Wooters \cite{W98} due to its
accessibility. Alternatively, the von Neumann entropy of macroscopic
(contiguous) subsystems can be used \cite{VPC04}.  A non-local measure
of entanglement was employed in the study of the
Affleck-Kennedy-Lieb-Tasaki (AKLT) model \cite{Vetal03,VMC04}.

In this paper, we will first discuss the effect of boundaries of the
Ising model on the entanglement properties using the concurrence as
measure of entanglement. We will then discuss a model which exhibits a
boundary phase transition, i.e., two Ising spins which are coupled to two
Kondo impurities. This model can be mapped onto the spin-boson model
and the concurrence can be computed which was originally defined for
the two Ising spins at the boundary \cite{StauberGuinea04}.  We
will then discuss various dissipative systems and compute the von
Neumann entropy, focusing the discussion on the cross-over from coherent
to incoherent oscillations \cite{Sta06}.  

The von Neumann entropy is a more general information measure than the
concurrence since the latter can only be defined for two spin-1/2
systems. The former can further be generalized to a measure which
relates non-contiguous sub-systems which shall be done in the third
part of this paper in the context of an harmonic chain. We note by
passing that the concurrence is an essentially local measure which
yields zero for all spin pairs which are not nearest or next-nearest
neighbors.

We close the introduction with some general remarks.
The models studied here, i.e., also the spin chains, 
can be interpreted as quantum systems characterized by a
small number of degrees of freedom coupled to a macroscopic reservoir.
These models show a crossover between different regimes, or even
exhibit a quantum critical point. As this behavior is induced by the
presence of a reservoir with a large number of degrees of freedom,
they can also be considered as a model of dephasing and loss of
quantum coherence. It is worth noting that there is a close connection
between models describing impurities coupled to a reservoir, and
strongly correlated systems near a quantum critical point, as
evidenced by Dynamical Mean Field Theory \cite{GKKR96}. In the limit
of large coordination, the properties of an homogeneous system can be
reduced to those of an impurity interacting with an appropriately
chosen reservoir. Hence, in the limit of large coordination the
entanglement between the quantum system and the reservoir near a phase
transition can be mapped onto the entanglement which develops in an
homogeneous system near a quantum critical point.

\section{Concurrence of the Ising model with general boundaries}
\subsection{The transverse Ising model}
We start with the homogeneous, one-dimensional transverse Ising model
with open boundary conditions and coupling parameter $\lambda$. The
two spins at the end are further connected by an additional coupling
parameter $\kappa$. For $\kappa=\lambda$, one recovers the Ising model
on a ring. The full Hamiltonian thus reads
\begin{align}
\label{IsingGen}
{\cal H}=-\lambda\sum_{i=1}^{N-1}\sigma_i^x\sigma_{i+1}^x-
\sum_{i=1}^{N}\sigma_i^z-\kappa\sigma_1^x\sigma_N^x\;,
\end{align}
where $\sigma_i^{x,z}$ are the $x,z$-components of the Pauli
matrices. 

To solve the model one first converts all the spin matrices into
spinless fermions $c_i$ with $\{c_i,c_{i'}\dag\}=\delta_{i,i'}$
\cite{LSM61,P70}.  This is done by performing a Jordan-Wigner
transformation
\begin{align}
\sigma_i^x&=\exp\left(i\pi\sum_{j=1}^{i-1}c_j\dag
c_j\right)(c_i+c_i\dag)\label{Jordan1}\;,\\ \sigma_i^z&=1-2c_i\dag
c_i\;.\label{Jordan2}
\end{align}
An additional Bogoliubov transformation then yields (up to a constant)
\begin{align}
{\cal H}=\sum_{n=1}^N\omega_n\eta_n\dag\eta_n\;,\quad\text{with}
\end{align}
\begin{align}
\label{eta}
\eta_n=\sum_{i=1}^N(g_{n,i}c_i+h_{n,i}c_i\dag)\;,
\end{align}
where $g_{n,i}$, $h_{n,i}$, and $\omega_n$ have to be determined
numerically for arbitrary ratio $\kappa/\lambda$. Due to the unitarity
of the Bogoliubov transformation, Eq. (\ref{eta}) is easily inverted
to yield
\begin{align}
\label{Bogoljubov}
c_i=\sum_{n=1}^N(g_{n,i}\eta_n+h_{n,i}\eta_n\dag)\quad.
\end{align}
For $\lambda=1$, the energy spectrum begins at zero energy which represents the critical point. For $\lambda>1$, apart from the extended states at finite energies there is also an additional zero-energy ``bound'' state. The emergence of the bound state can be interpreted as a loss of coherence. Since it is connected to the appearance of a zero energy mode which is inherent to quantum phase transitions, we believe that this loss of coherence is a general feature that provokes the change in entanglement and that this view can be generalized to other systems with quantum phase transitions. For more details, see appendix \ref{App:Bogoljubov}.

\subsection{Concurrence as information measure}
We are interested in the reduced density matrix $\rho(i,j)$
represented in the basis of the eigenstates of $\sigma_z$. It is
formally obtained from the ground-state wave function after having
integrated out all spins but the ones at position $i$ and $j$. As
measure of entanglement, we use the concurrence between the two
spins, ${\cal C}(\rho(i,j))$. It is defined as
\begin{align}
\label{Cdefinition}
{\cal C}(\rho(i,j))=\text{max}\{0,\lambda_1-\lambda_2-\lambda_3-\lambda_4\}
\end{align}
where the $\lambda_i$ are the (positive) square roots of the
eigenvalues of $R=\rho\tilde\rho$ in descending order. The spin
flipped density matrix is defined as
$\tilde\rho=\sigma_y\otimes\sigma_y\rho^*\sigma_y\otimes\sigma_y$,
where the complex conjugate $\rho^*$ is again taken in the basis
of eigenstates of $\sigma_z$. It will be instructive to also
consider the ``generalized concurrence''
\begin{align}
\label{C*definition}
{\cal C}^*(\rho(i,j))=\lambda_1-\lambda_2-\lambda_3-\lambda_4\quad.
\end{align}

The reduced density matrix $\rho(i,j)\to\rho$ (from now on we
drop the indices $i$ and $j$) can be related to correlation
functions. For this, we write the ground-state wave function as
the superposition of the four states
\begin{align}
|\psi_0\rangle=|\uparrow\uparrow\rangle|\phi_{\uparrow\uparrow}\rangle+
|\uparrow\downarrow\rangle|\phi_{\uparrow\downarrow}\rangle+|\downarrow\uparrow\rangle|
\phi_{\uparrow\downarrow}\rangle+|\downarrow\downarrow\rangle|\phi_{\downarrow\downarrow}\rangle,
\end{align}
where the first ket denotes the state of the two spins at position $i$
and $j$ and the second ket the corresponding state of the rest of the spin
system. The matrix element
$\rho_{\uparrow\uparrow,\downarrow\downarrow}=
\langle\phi_{\uparrow\uparrow}|\phi_{\downarrow\downarrow}\rangle$,
e.g., is thus given by
$\rho_{\uparrow\uparrow,\downarrow\downarrow}=\langle\sigma_i^+\sigma_j^+\rangle$,
where $\sigma_i^\pm=(\sigma_i^x\pm\sigma_i^y)/2$.

Due to the invariance of the Hamiltonian under
$\sigma_i^x=-\sigma_i^x$, at least eight components of the reduced
density matrix are zero (for finite $N$). The diagonal entries
read:
\begin{align}
\rho_1\equiv\rho_{\uparrow\uparrow,\uparrow\uparrow}&=(1+
\langle\sigma_i^z\rangle+\langle\sigma_j^z\rangle+\langle\sigma_i^z\sigma_j^z\rangle)/4\\
\rho_2\equiv\rho_{\uparrow\downarrow,\uparrow\downarrow}&=(1-
\langle\sigma_i^z\rangle+\langle\sigma_j^z\rangle-\langle\sigma_i^z\sigma_j^z\rangle)/4\\
\rho_3\equiv\rho_{\downarrow\uparrow,\downarrow\uparrow}
&=(1+\langle\sigma_i^z\rangle-\langle\sigma_j^z\rangle-
\langle\sigma_i^z\sigma_j^z\rangle)/4\\
\rho_4\equiv\rho_{\downarrow\downarrow,\downarrow\downarrow}
&=(1-\langle\sigma_i^z\rangle-\langle\sigma_j^z\rangle+\langle\sigma_i^z\sigma_j^z\rangle)/4
\end{align}
The non-zero off-diagonal entries are
\begin{align}
\rho_+\equiv\rho_{\uparrow\uparrow,\downarrow\downarrow}&=\langle\sigma_i^+\sigma_j^+\rangle\\
\rho_-\equiv\rho_{\uparrow\downarrow,\downarrow\uparrow}&=\langle\sigma_i^+\sigma_j^-\rangle\quad.
\end{align}
The positive square roots of the eigenvalues of $R$ are then given
by $|\sqrt{\rho_1\rho_4}\pm\rho_+|$ and
$|\sqrt{\rho_2\rho_3}\pm\rho_-|$. Due to the semi-definiteness of
the density matrix $\rho$, we can drop the absolute values, i.e.,
$\sqrt{\rho_1\rho_4}\pm\rho_+$ and $\sqrt{\rho_2\rho_3}\pm\rho_-$.

We now define
$I_1\equiv\rho_1\rho_4-\rho_2\rho_3=4(\langle\sigma_i^z\sigma_{j}^z
\rangle-\langle\sigma_i^z\rangle\langle\sigma_j^z\rangle)$ and
$I_2\equiv\rho_+^2
-\rho_-^2=-\langle\sigma_i^x\sigma_{j}^x\rangle\langle\sigma_i^y\sigma_{j}^y\rangle$.
For a homogeneous model, we have $I_1\geq0$ and
$I_2\geq0$.\cite{P70} The largest eigenvalue of Eq.
(\ref{Cdefinition}) is thus given by
$\lambda_1=\sqrt{\rho_1\rho_4}+|\rho_+|$ and the concurrence reads
\begin{align}
{\cal C}^*(i,j)&=2(|\rho_+|-\sqrt{\rho_2\rho_3})\quad.
\end{align}
We note that the above expression also holds for the generalized
boundary conditions. For a homogeneous system, it can be further
simplified to
\begin{align}
{\cal C}^*(i,j)&=({\cal O}_{i,j}-1)/2
\end{align}
where we introduced the total order ${\cal O}_{i,j}\equiv\sum_{\alpha=x,y,z}|
\langle\sigma_i^\alpha\sigma_{j}^\alpha\rangle|$.

\subsection{Numerical results}
\subsubsection{Open boundary conditions}
\begin{figure}[t]
  \begin{center}
    \includegraphics*[width=3in,angle=0]{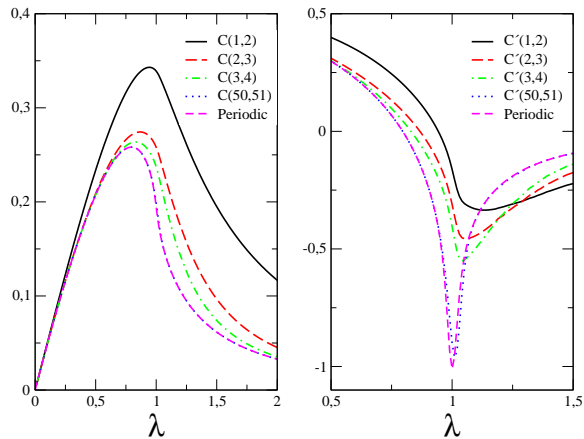}
    \caption{Left hand side: The nearest neighbor concurrence
    of the open boundary Ising model for different locations relative
    to the end as function of $\lambda$ for $N=101$. Right hand side:
    The derivative of the concurrence with respect to $\lambda$.}
    \label{Boundary}
\end{center}
\end{figure}
\begin{figure}[t]
  \begin{center}
    \includegraphics*[width=3in,angle=0]{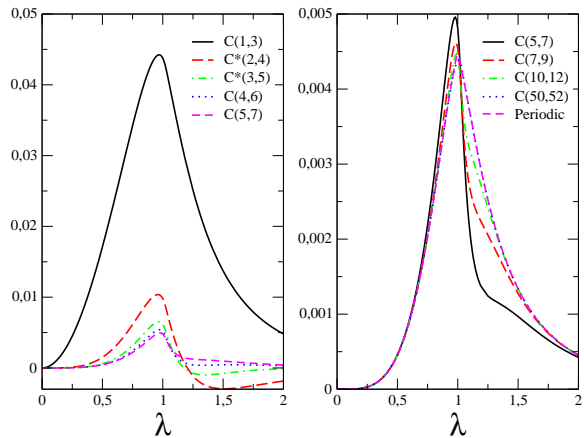}
    \caption{The generalized next-nearest neighbor concurrence of the open
    boundary Ising model for different locations relative to the
    end of the chain as function of $\lambda$ for $N=101$.}
    \label{NextNearest}
\end{center}
\end{figure}
We first consider the nearest neighbor concurrence of the Ising
chain with open boundaries ($\kappa=0$) for a fixed number of
sites $N=101$ as parameter of $\lambda$, but for various positions
relative to the end of the chain. The results are displayed on the
left hand side of Fig. \ref{Boundary}. As expected, the
concurrence of the periodic model is approached as one moves
inside the chain and the difference between ${\cal C}(50,51)$ and
${\cal C}(i,i+1)$ of the periodic system is hardly seen.
Nevertheless, the derivative of the concurrence with respect to
the coupling parameter $\lambda$, ${\cal C}'\equiv d{\cal
C}/d\lambda$, still shows appreciable differences for
$\lambda\approx1$ (right hand side of Fig. \ref{Boundary}).

We also investigated the scaling behavior of the minimum of ${\cal
C}'(1,2)$, $\lambda_{min}$, for different systems sizes up to
$N=231$. We did not find finite-size scaling behavior for the
position of the minimum as is the case for the translationally
invariant model\cite{Oetal02}. The curve of ${\cal C}'(1,2)$,
shown on the right hand side of Fig. \ref{Boundary}, is thus
already close to the curve for $N\to\infty$ with a broad minimum
around $\lambda_{min}\approx1.1$.

The absence of finite-size scaling of the concurrence is also
manifested in the case of the next-nearest neighbor concurrence
for different system sizes $N$. Whereas for the periodic system
the maximum of ${\cal C}(i,i+2)$ decreases monotonically for
$N\to\infty$,\cite{Oetal02} there is practically no change of
${\cal C}(1,3)$ of the open chain for $N\gapp51$.

In Fig. \ref{NextNearest}, the generalized next-nearest neighbor
concurrence ${\cal C}^*(i,i+2)$ of the open boundary Ising model
is shown for different locations relative to the end of the chain
as function of $\lambda$ for $N=101$. On the left hand side of
Fig. \ref{NextNearest}, results are shown for sites close to the
end of the chain. Notice that the generalized concurrence becomes
negative for $i=2,3$ for $\lambda>1$ which is not related to the
quantum phase transition. The crossover of the boundary behavior
to the bulk behavior is thus discontinuous. On the right hand side
of Fig. \ref{NextNearest}, the next-nearest neighbor concurrence
approaches the result of the system with periodic boundary
conditions as one moves inside the chain.

\subsubsection{Generalized boundary conditions}
\begin{figure}[t]
  \begin{center}
    \includegraphics*[width=3in,angle=0]{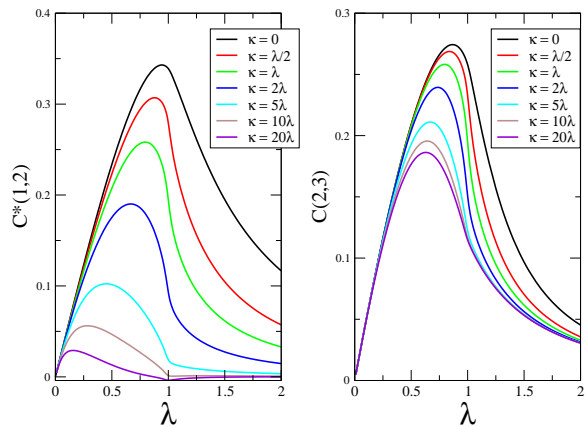}
    \caption{Left hand side: The generalized nearest neighbor concurrence of
    the closed Ising chain for various coupling strengths $\kappa$ as function
    of $\lambda$. Left hand side: ${\cal C}^*(1,2)$. Right hand side: ${\cal C}(2,3)$.}
    \label{Kopplung}
\end{center}
\end{figure}
\begin{figure}[t]
  \begin{center}
    \includegraphics*[width=3in,angle=0]{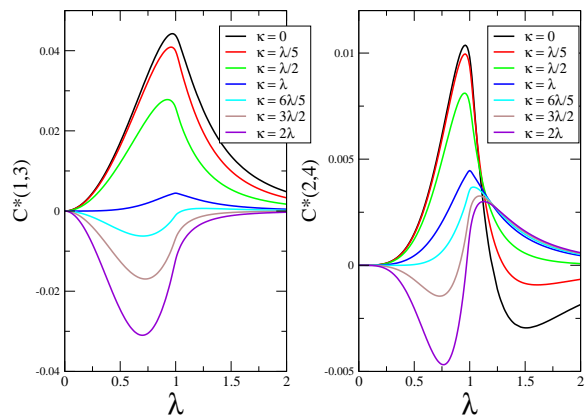}
    \caption{The generalized next-nearest neighbor concurrence of the closed Ising chain for
    various coupling strengths $\kappa$ as function of $\lambda$. Left hand side:
    ${\cal C}^*(1,3)$. Right hand side: ${\cal C}^*(2,4)$.}
    \label{Kopplung2}
\end{center}
\end{figure}
We now discuss the concurrence for the generalized boundary
conditions, introducing the parameter $\kappa$. On the left hand
side of Fig. \ref{Kopplung}, the generalized concurrence of the
first two spins ${\cal C}^*(1,2)$ is shown as function of
$\lambda$ for various coupling strengths $\kappa=0,..,20\lambda$
and $N=101$. For increasing $\kappa>0$, the curves indicate
stronger non-analyticity at $\lambda\approx1$. For $\kappa\gapp
20\lambda$, the generalized concurrence becomes negative around
$\lambda=1$ and is ''significantly'' positive only in the quantum
limit of a strong transverse field ($\lambda<1$). A similar
behavior of the concurrence is also found in the case of finite
temperatures.\cite{Arnesen01,ON02}

On the right hand side of Fig. \ref{Kopplung}, the concurrence of
the second two spins ${\cal C}(2,3)$ is shown. All curves display
similar behavior. There is thus a rapid crossover from the
boundary to the bulk-regime and the concurrence of periodic
boundary conditions is approached for all $\kappa$ as one moves
further inside the chain.

To close, we discuss the next-nearest neighbor concurrence ${\cal
C}(i,i+2)$ for various values of $\kappa$ and $N=101$. On the left
hand side of Fig. \ref{Kopplung2}, the generalized concurrence of
the first and the third spin, ${\cal C}^*(1,3)$, is shown. For
$\kappa\lapp1$, ${\cal C}^*(1,3)$ is positive for all $\lambda$.
For $\kappa\gapp1$, ${\cal C}^*(1,3)$ first becomes negative for
$\lambda<1$. For $\kappa\gapp1.5$ ${\cal C}^*(1,3)$ is negative
for all $\lambda$. On the right hand side of Fig. \ref{Kopplung2},
the generalized concurrence of the second and the forth spin,
${\cal C}^*(2,4)$, is shown. For $\kappa\lapp\lambda/2$, the
${\cal C}^*(2,4)$ is negative for $\lambda\gapp1$. For
$\kappa\gapp3\lambda/2$, the ${\cal C}^*(2,4)$ is negative for
$\lambda\lapp1$. Nevertheless, the maximum value is close to
$\lambda=1$ for all cases.

We finally note that the third neighbor concurrence remains zero for all $\lambda$ and all $\kappa$.

\subsection{Summary}
To conclude, we have calculated the entanglement between qubits at the
boundary of a spin chain, whose parameters are tuned to be near a
quantum critical point. The calculations show a behavior which differs
significantly from the that inside the bulk of the chain.  Although
the spins are part of the critical chain, we find no signs of the
scaling behavior which can be found in the bulk.  Still, we could
identify a boundary regime, basically given by the first site, and a
crossover regime of approximately 10 sites till the bulk behavior is
reached. We use the same approach as done previously for the
bulk\cite{Oetal02,ON02}, although it should be noted that the
existence of a finite order parameter in the ordered phase will change
these results if the calculations were performed in the presence of an
infinitesimal applied field.
\section{Concurrence at a boundary phase transition}
\label{BoundaryPT}
In order to observe critical behavior of the concurrence at the
boundary, one has to consider a different model than the simple
transverse Ising chain. One possibility would be to introduce an isotropic
coupling from spin $N$ to spin $1$ which would lead to an interaction
term containing four fermionic operators. A simple solution is
thus not possible anymore. In the following, we will consider a
similar model, but which can easily be mapped onto the spin-boson
model.

\subsection{The model}
The model with isotropic coupling between the two spins at the end is
similar to the model introduced by Garst {\it et al.} \cite{Getal03}
(see also Ref. \cite{VBH02}). It describes two spin-1/2 systems
attached to two different electronic reservoirs. They further interact
among themselves through an Ising term. We can write the Hamiltonian
as
\begin{eqnarray}
H &= &{H}_{K1} + {H}_{K2} + I S_{z1} S_{z2}\;,
\nonumber \\
{H}_{Ki} &= &\sum_k  \epsilon_{k , \mu} c^{\dag}_{k,\mu ,i}
c_{k, \mu , i} + J \sum_{k , k' , \mu , \nu} c_{k , \mu , i}
\vec{\sigma}_{\mu , \nu} c_{k', \nu , i} \vec{S}_i \nonumber\;. \\ &
& \label{hamil}
\end{eqnarray}
To evaluate the concurrence of the two spins, the $4
\times 4$ reduced density matrix in the basis of the eigenstates of 
$S_{z1}$ and $S_{z2}$ is needed.

The system described by Eq.(\ref{hamil}) undergoes a
Kosterlitz-Thouless transition between a phase with a doubly
degenerate ground state and a phase with a non degenerate ground
state. This transition is equivalent to that in the dissipative
two-level system\cite{Letal87,W99} as function of the strength of
the dissipation. We define the dissipative two-level system as
\begin{equation}
{H}_{TLS} = \Delta \sigma_x + \sum_k | k | b^{\dag}_k b_k +
\lambda \sigma_z \sum_k \sqrt{k} ( b^{\dag}_k + b_k )\;. \label{TLS}
\end{equation}
The strength of the dissipation can be characterized by a
dimensionless parameter, $\alpha \propto \lambda^2$, and the model
undergoes a transition for $\tilde{\Delta} = \delta / \omega_c \ll
1$, where $\omega_c$ is the cutoff, and $\alpha = 1$. The Kondo
model can be mapped onto this model by taking $\Delta
\propto \tilde{J_\perp}$ and $1 - \alpha \propto \tilde{J_z}$ \cite{GHM95}.

To understand the equivalence between these two models, it is best
to to consider the limit $I/J \gg 1$ (the transition takes place
for all values of this ratio). Let us suppose that $I > 0$ so that
the Ising coupling is antiferromagnetic. The Hilbert space of the
two impurities has four states. The combinations $| \uparrow
\uparrow \rangle$ and $| \downarrow  \downarrow \rangle$ are
almost decoupled from the low energy states, and the transition
can be analyzed by considering only the $| \uparrow \downarrow
\rangle $ and $ | \downarrow \uparrow \rangle$ combinations. Thus,
we obtain an effective two state system. The transition is driven
by the spin flip processes described by the Kondo terms. These
processes involve two simultaneous spin flips in the two
reservoirs. Hence, the operator which induces these spin flips
leads to the correspondence $\tilde{\Delta} \leftrightarrow 
J_\perp^2 / ( I \omega_c )$. The scaling dimension of this term,
in the Renormalization Group sense, is reduced with respect to the
ordinary Kondo Hamiltonian, as two electron-hole pairs must be
created. This implies the equivalence $2 - \alpha \leftrightarrow
\tilde{J_z}$. Hence, the transition, which for the ordinary Kondo
system takes place when changing the sign of $J_z$ now requires a
finite value of $J_z$.
\subsection{Calculation of the concurrence}
\label{ConcSB}
The $4 \times 4$ reduced density matrix can be decomposed into a
$2 \times 2$ box involving the states $| \uparrow \downarrow
\rangle$ and $| \downarrow \uparrow \rangle$, which contains the
matrix elements which are affected by the transition, and the
remaining elements involving $| \uparrow \uparrow \rangle$ and $|
\downarrow \downarrow \rangle$ which are small, and are not
modified significantly by the transition. Neglecting these
couplings, we find that two of the four eigenvalues of the density
matrix are zero. The other two are determined by the matrix
\begin{equation}
\tilde{\rho} \equiv \frac{1}{2}\left( \begin{array}{cc} 1 + \langle
\sigma_z \rangle &\langle \sigma_x \rangle \\ \langle \sigma_x
\rangle &1 - \langle \sigma_z \rangle\;. \end{array}
\right) \label{matrix}
\end{equation}
where the operator $\tilde{\sigma}$ is defined using the standard
notation of the dissipative two level system, Eq. (\ref{TLS}). The
entanglement can be written as
\begin{equation}
{\cal C} = \sqrt{ \langle \sigma_z \rangle^2 + \langle \sigma_x
\rangle^2}\;. \label{entanglement_TLS}
\end{equation}
The value of $\langle \sigma_z \rangle$ is the order parameter of
the transition. The value of $\langle \sigma_x \rangle$, at zero
temperature, can be calculated from
\begin{equation}
\langle \sigma_x \rangle = \frac{\partial E_0}{\partial \Delta}\;.
\label{der_ener}
\end{equation}
where $E_0$ is the energy of the ground state. Using scaling 
arguments (see appendix \ref{App:Scaling}), it can be written as followed:
\begin{equation}
E_0 = \left\{ \begin{array}{lr} \frac{C}{1 - 2
\alpha} \left[ \Delta \left( \frac{\Delta}{\omega_c}
\right)^{\frac{\alpha}{1 - \alpha}} - \frac{\Delta^2}{\omega_c}
\right] &0 < \alpha < \frac{1}{2} \\ 2 C
\frac{\Delta^2}{\omega_c} \log \left( \frac{\omega_c}{\Delta}
\right) &\alpha = \frac{1}{2}
\\ \frac{C}{2 \alpha - 1} \left[ \frac{\Delta^2}{\omega_c} - \Delta
\left( \frac{\Delta}{\omega_c} \right)^{\frac{\alpha}{1 - \alpha}}
\right] &\frac{1}{2} < \alpha < 1 \\ C \left(
\frac{\Delta^2}{\omega_c} - C' \omega_c e^{- \frac{C''
\omega_c}{\Delta}} \right) &\alpha \sim 1 \\ C
\frac{\Delta^2}{\omega_c} & \alpha > 1
\end{array} \right. \label{ener_TLS}
\end{equation}
where $C , C'$ and $C''$ are numerical constants.

If the density matrix is calculated in the absence of a symmetry
breaking field, $\langle \sigma_z \rangle = 0$ even in the ordered
phase. Then, from Eq.(\ref{entanglement_TLS}), the concurrence is
given by ${\cal C} = |\langle \sigma_x \rangle|$, which is
completely determined using Eqs. (\ref{der_ener}) and
(\ref{ener_TLS}). In the limit $\Delta / \omega_c \ll 1$ the
interaction with the environment strongly suppresses the
entanglement. We expect unusual behavior of the concurrence for
$\alpha = 1 / 2$ and $\alpha = 1$. The point $\alpha = 1/2$ marks
the loss of coherent oscillations between the two
states\cite{G85,note2}, although the ground state remains non
degenerate. Following the analysis in \cite{Oetal02}, we analyze
the behavior of $\partial {\cal C} /
\partial \alpha$, as $\alpha$ is the parameter which determines
the position of the critical point. The strongest change of
this quantity occurs for $\alpha = 1 / 2$, where:
\begin{equation}
\left. \frac{\partial {\cal C}}{\partial \alpha} \right|_{\alpha =
1/2} \sim \frac{\Delta}{\omega_c} \log \left(
\frac{\omega_c}{\Delta} \right)
\end{equation}
On the other hand, near $\alpha = 1$ the value of $\partial {\cal C} /
\partial \alpha$ is continuous, as the influence of the critical point
has a functional dependence, when $\alpha \rightarrow \alpha_c$, of
the type $e^{- c / ( \alpha_c - \alpha )}$. This is the standard
behavior at a Kosterlitz-Thouless phase transition. This result
suggest that the entanglement is more closely related to the presence
of coherence between the two qubits than to the phase transition. The
transition takes place well after the coherent oscillations between
the $| \uparrow \downarrow \rangle$ and $| \downarrow \uparrow
\rangle$ states are completely suppressed. We note though that with a
symmetry breaking field, there is a discontinuity of the concurrence at
the phase transition \cite{Kopp07}.

\section{Von Neumann entropy for dissipative systems}

In this section, we will use the von Neumann entropy as measure of entanglement. It is defined for any bipartite system with a ground-state $|\psi\rangle$ by introducing the reduced density matrix with respect to one of the subsystem. For the two subsystems $A$ and $B$, it reads 
\begin{align}
E(\psi)=-\text{Tr}(\rho_A\ln\rho_A)\quad,\quad\rho_A=\text{Tr}_B(|\psi\rangle\langle\psi|)\;.
\label{entropy}
\end{align}
In contrast to the concurrence, an analytic expression of the reduced
density matrix $\rho_A$ does not automatically lead to an analytic
expression for the von Neumann entropy. In the following, we show that
in the case of integrable dissipative models, the von Neumann entropy
can be obtained analytically. We then also discuss the von Neumann
entropy of the spin-boson model.
\subsection{Integrable quantum dissipative systems}
Modeling the environment by a set of harmonic oscillators \cite{Cal83}, the canonical (integrable) model for dissipative systems is described by the following Hamiltonian:
\begin{align}
\label{HOscillator}
    H&=\frac{p^2}{2}+\frac{\omega_0^2}{2} q^2+
     \sum_{\alpha}\Big(\frac{p_{\alpha}^2}{2}
        +\frac{1}{2}
        \omega_{\alpha}^2\big(x_{\alpha}-\frac{\lambda_{\alpha}}{\omega_{\alpha}^2}q\big)^2\Big)
\end{align}
The operators obey the canonical commutation relations which
read ($\hbar=1$)
\begin{align}
 \left[q,p\right]=i\quad,\quad\left[x_{\alpha},p_{\alpha^{\prime}}\right]=i
 \delta_{\alpha,\alpha'}\quad.
\end{align}
The coupling of the system to the bath is completely determined by the spectral function
\begin{align}
\label{SpectralJHO}
J_{HO}(\omega)=\frac{\pi}{2}\sum_\alpha\frac{\lambda_\alpha^2}{\omega_\alpha}\delta(\omega-\omega_\alpha).
\end{align}
In the following, we will consider a Ohmic bath with $J_{HO}(\omega)=\eta\omega$ for $\omega\ll\omega_c$ and $J_{HO}(\omega)=0$ for $\omega\gg\omega_c$, $\omega_c$ being the cutoff frequency.
\subsubsection{Caldeira-Leggett model}
Let us first consider the free dissipative particle, i.e., we set $\omega_0=0$. The model was introduced by Caldeira and Leggett \cite{Cal83b} and further investigated by Hakim and Ambegaokar \cite{Hak85}. The latter authors obtained the reduced density matrix via diagonalization of the Hamiltonian. In real space, it reads
\begin{align}
\langle x|\rho_A|x'\rangle=e^{-a(x-x')^2}/L\quad,\quad a=\frac{1}{4}\frac{\eta}{\pi}\ln\left(1+\frac{\omega_c^2}{\eta^2}\right)\;
\label{densitymatrix}
\end{align}
where $\eta$ denotes the phenomenological friction coefficient and $\omega_c$ is the cutoff frequency of the bath, introduced below Eq. (\ref{SpectralJHO}). Furthermore, $L\to\infty$ denotes the system size and in contrast to the use of Eq. \ref{densitymatrix} in Ref. \cite{Hak85}, here the normalization is crucial to assure Tr$\rho_A=1$. 

In order to calculate the entropy of the system, we Taylor expand the logarithm
\begin{align}
\label{LogEx}
\ln\rho_A=-\sum_{n=1}\frac{(1-\rho_A)^n}{n}=-\sum_{n=1}\frac{1}{n}\sum_{k=0}^n{n\choose k}(-1)^k\rho_A^k\;.
\end{align} 
Further we have
\begin{align}
\langle x|\rho_A^k|x'\rangle=\sqrt{\frac{\pi}{a}}^{k-1}\sqrt{\frac{1}{k}}e^{-\frac{a}{k}(x-x')^2}/L^k
\end{align}
proved by induction. With the identity
\begin{align}
\sqrt{\frac{1}{k}}=\frac{1}{\sqrt{\pi}}\int dxe^{-kx^2}
\end{align}
we thus obtain for the specific entropy (for general dimension $d$)
\begin{align}
S=\frac{d}{2}\left[\ln(aL^2)+\ln(e\pi)\right].
\end{align}
Comparing the above result with the entropy of a particle in a canonical ensemble, we identify $a\sim\lambda^{-2}\propto T$ with $\lambda$ denoting the thermal de Broglie wavelength and $T$ the temperature of the canonical ensemble. Notice that the entropy of a free dissipative particle shows no non-analyticity.
\subsubsection{Dissipative harmonic oscillator}
We now include the harmonic potential, i.e., $\omega_0\neq0$.
The reduced density matrix of the damped harmonic oscillator is given by \cite{W99}
\begin{align}
\langle x|\rho_A&|x'\rangle=\sqrt{\frac{4b}{\pi}}e^{-a(x-x')^2-b(x+x')^2}\;,
\end{align}
with $a=\frac{\langle p^2\rangle}{2}$ and $b=\frac{1}{8\langle q^2\rangle}$. The above expression is deduced such that the correct variance for position and momentum is obtained. At $T=0$ the expectation values are given by
\begin{align}
\langle q^2\rangle&=\frac{1}{2\omega_0}f(\kappa)\quad,\quad
\langle p^2\rangle=\omega_0^2(1-2\kappa^2)\langle q^2\rangle+\frac{2\omega_0\kappa}{\pi}\ln\left(\frac{\omega_c}{\omega_0}\right)\;,
\end{align} 
with $\kappa=\eta/2\omega_0$ and 
\begin{align}
f(\kappa)=\frac{1}{\pi}\frac{\ln\left[(\kappa+\sqrt{\kappa^2-1})/(\kappa-\sqrt{\kappa^2-1})\right]}{\sqrt{\kappa^2-1}}\quad.
\end{align}
The parameter $\kappa$ represents the friction parameter and the system experiences a crossover from coherent to incoherent oscillations at $\kappa=1$.

Taylor expanding the logarithm of the entropy, Eq. (\ref{LogEx}), leads to the evaluation of the general $n$-dimensional integral
\begin{align}
\int_{-\infty}^{\infty}dx_1..dx_n\exp\left(-\sum_{i,j=1}^nx_iA_{i,j}x_j\right)=\frac{\pi^{n/2}}{\sqrt{\text{det}A}}
\end{align}
where $A$ is given by the translationally invariant tight-binding matrix with $A_{i,i}=2(a+b)$, $A_{i+1,i}=A_{i,i+1}=-(a-b)$ ($n+1\equiv1$) and zero otherwise. The determinant of the matrix is given by its eigenvalues and reads
\begin{align}
\text{det}A=(2a)^n(1-b/a)^n\prod_{m=1}^n\left[1+\frac{2b}{a-b}-\cos k_m\right]
\end{align}
with $k_m=2\pi m/n$. 

The determinant can be easily evaluated for large cut-offs $\omega_c\rightarrow\infty$ \cite{Sta06}: Considering the $n$-dimensional translationally invariant, but non-Hermitian matrix $\widetilde A_{i,i}=1$, $\widetilde A_{i+1,i}=1-\varepsilon$ ($n+1\equiv1$) and zero otherwise, one obtains the following formula:
\begin{align}
\prod_{m=1}^n\left[1+\frac{\varepsilon^2}{2(1-\varepsilon)}-\cos k_m\right]=\frac{(1-(1-\varepsilon)^{n})^2}{2^n(1-\varepsilon)^n}
\end{align}
For $\omega_c/\omega_0\gg1$, we have
\begin{align}
a/b&=4\langle q^2\rangle\langle p^2\rangle\notag\\
&=f(\kappa)\left[(1-2\kappa^2)f(\kappa)+\frac{4\kappa}{\pi}\ln\left(\frac{\omega_c}{\omega_0}\right)\right]\gg1.
\end{align}
In this limit, we can thus set $\varepsilon^2=4b/a\ll1$ and the $n$-dimensional integral can be approximated to yield
\begin{align}
\int dx\langle x|\rho_A^n|x\rangle&\rightarrow\frac{\tilde\varepsilon^n}{1-(1-\varepsilon)^n}\;,
\end{align}
with $\tilde\varepsilon\equiv\varepsilon\sqrt{1-\varepsilon}/\sqrt{1-\varepsilon^2/4}$.
Expanding the denominator as geometrical series,
we then have for the entropy
\begin{align}
S&=-\left(\frac{\tilde\varepsilon}{\varepsilon}\ln\tilde\varepsilon+\frac{\tilde\varepsilon}{\varepsilon^2}\ln(1-\varepsilon)\right).
\label{entropieO}
\end{align}
In the limit $\varepsilon\approx\tilde\varepsilon\ll1$, the leading behavior of the entropy is given by $S\sim\ln(a/b)$. 

The determinant can also be calculated exactly as was done in Ref. \cite{Kopp07}. This yields the exact expression of the von Neumann entropy,
\begin{align}
S=-\frac{1}{2}\ln\left(\frac{4b}{a-b}\right)-\frac{a}{2b}\ln\left(\frac{\sqrt{a}-\sqrt{b}}{\sqrt{a}+\sqrt{b}}\right)\;.
\label{EntropieHarmOsci}
\end{align}

\begin{figure}[t]
  \begin{center}
    \includegraphics*[width=3.5in,angle=-90]{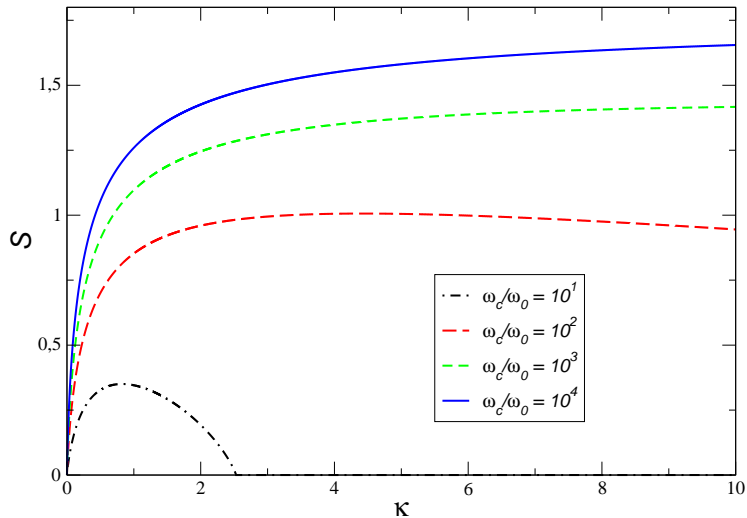}
    \caption{The entropy $S$ of the dissipative oscillator with Ohmic coupling as function of the dimensionless coupling strength $\kappa$ for various cut-off frequencies $\omega_c$.}
    \label{EntropieOsci}
\end{center}
\end{figure}

The von Neumann entropy is plotted in Fig. \ref{EntropieOsci} for various cutoff energies $\omega_c$ as function of the coupling constant $\kappa$. Notice that in all curves a crossover behavior occurs at $\kappa\approx1$, where coherent and incoherent oscillations interchange. 
\subsection{Spin-boson model}
In section \ref{BoundaryPT}, the spin-boson model or dissipative two-level system was already introduced and the concurrence was calculated in the context of a two-impurity Kondo model. Here, we want to compute the von-Neumann entropy for this system. Since we will discuss several bath types, the Hamiltonian without bias shall be defined with general coupling constants $\lambda_k$ as
\begin{align}
    H=\frac{\De_0}{2}\sigma_x  
        +\sum_k\omega_{k}b_{k}\dag b_k 
        +\sigma_z\sum_k\frac{\lambda_{k}}{2}(b_k+b_k\dag)\quad.
\end{align}
Again, the operators $b_{k}^{(\dagger)}$ resemble the bath degrees of freedom
and $\sigma_x$, $\sigma_y$, $\sigma_z$ denote the Pauli spin matrices. The coupling constants $\lambda_k$ give rise to the spectral function
\begin{align}
J(\omega)=\sum_k\lambda_k^2\delta(\omega-\omega_k).
\label{spectralFunctionSB}
\end{align}

In the relevant low-energy regime, the spectral function is generally parameterized as a power-law, i.e., $J(\omega)\propto2\alpha\omega^s\Lambda_0^{1-s}$ where $\alpha$ denotes the coupling constant, $s$ the bath type ($s=1$ defines the previously discussed Ohmic dissipation) and  $\Lambda_0$ the cutoff-frequency. The change in notation ($\omega_c\rightarrow\Lambda_0$) will be convenient in the context of the scaling approach.

With $A$ denoting the spin-1/2 system, the reduced density matrix of the spin-boson model is given by
\begin{align}
\rho_A=\frac{1}{2}
\begin{pmatrix}
1 + \langle
\sigma_z \rangle &\langle \sigma_x \rangle \\ \langle \sigma_x
\rangle &1 - \langle \sigma_z \rangle
\end{pmatrix}.
\label{entropySB}
\end{align}
Since there is no symmetry breaking field in the above Hamiltonian, we set $\langle \sigma_z\rangle=0$. The eigenvalues are thus given by $\lambda_\pm=(1\pm\langle \sigma_x \rangle)/2$ and the entropy reads
\begin{align}
\label{Ssb}
S=-\frac{1}{2}\left[\ln\left((1-\langle \sigma_x \rangle^2\right)/4)+\langle \sigma_x \rangle\ln\left(\frac{1+\langle \sigma_x \rangle}{1-\langle \sigma_x \rangle}\right)\right].
\end{align}
The value of $\langle \sigma_x \rangle$, at zero
temperature, is given by 
\begin{equation}
\langle \sigma_x \rangle = 2\frac{\partial E_0}{\partial \Delta_0}
\end{equation}
where $E_0$ is the energy of the ground-state. To obtain the ground-state energy, a scaling analysis for the free energy at arbitrary temperature is considered as before (see appendix \ref{App:Scaling}). $E_0( \Delta_0 )$ and $\langle \sigma_x \rangle$ will then set the basis for our discussion on the entanglement properties of the spin-boson model, see Eq. (\ref{Ssb}).

\subsubsection{Ohmic dissipation}
In the Ohmic case ($s=1$), there is a phase transition at zero temperature at the critical coupling strength $\alpha=1+O(\Delta/\Lambda_0)$ \cite{Bra82,Cha82}. The transition is also reflected by the renormalized tunnel matrix element $\Dren$ which reads $\Dren=\De_0(\De_0/\Lambda_0)^{\alpha/(1-\alpha)}$ for $\alpha<1$ and $\Dren=0$ for $\alpha>1$. 

\begin{figure}[t]
  \begin{center}
    \includegraphics*[width=3.5in,angle=0]{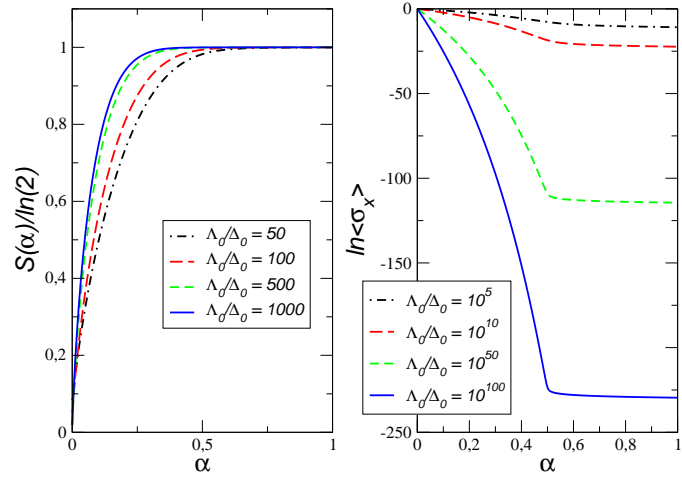}
    \caption{Left hand side: The entropy $S$ of the spin-boson model with Ohmic coupling in units of $\ln(2)$ as function of the dimensionless coupling strength $\alpha$ for various cut-off frequencies $\Lambda_0$. The right hand side shows $\ln\langle \sigma_x\rangle$ which derivative with respect to the coupling approaches a step-like function in the scaling limit $\Delta_0/\Lambda_0\rightarrow0$.}
    \label{EntropieSB}
\end{center}
\end{figure}

The von Neumann entropy of the spin-boson model with Ohmic dissipation was first discussed by means of a renormalization group approach\cite{CMcK03} and later also by the thermodynamical Bethe ansatz\cite{KoHur07}. Here, we will obtain the von Neumann entropy within a scaling approach which can also be extended to non-Ohmic dissipation. In this approach, the free energy is given by (see appendix \ref{App:Scaling})
\begin{equation}
F = \int_{\Dren}^{\Lambda_0} \left( \frac{\Delta ( \Lambda )}{\Lambda}
\right)^2 d \Lambda\;.
\label{int_ener}
\end{equation}
With $\Delta(\Lambda)=\Delta_0(\Lambda/\Lambda_0)^\alpha$, the ground state energy $E_0$ is then given by Eq.(\ref{ener_TLS}) and the discussion is similar to the one in section \ref{ConcSB}. For $\alpha=1/2$, we thus have
\begin{equation}
\frac{d\ln\langle\sigma_x\rangle}{d\alpha}\Big|_{\alpha=1/2}\propto\ln\left(\frac{\Delta_0}{\Lambda_0}\right)\;.
\label{Divergence}
\end{equation}
In the scaling limit $\Delta_0/\Lambda_0\rightarrow\infty$, this quantity diverges logarithmically. This is shown on the right hand side of Fig. \ref{EntropieSB}. The entropy $S$ of the dissipative two-level system with Ohmic coupling is plotted in Fig. \ref{EntropieSB} as function of the dimensionless coupling strength $\alpha$ for various cutoff frequencies $\Lambda_0$. The entropy quickly saturates after the transition from coherent to incoherent oscillations at $\alpha=1/2$ as can be seen in terms of $\ln\langle\sigma_x\rangle$ on the left hand side of Fig. \ref{EntropieSB}.

\subsubsection{Non-Ohmic dissipation}
The calculation of $E_0 ( \Delta_0 )$ and $\langle \sigma_x \rangle$ can be
extended to the spin-boson model with non-Ohmic dissipation ($s\neq1$). In general, the dependence of the effective tunneling term on the cutoff, $\Delta( \Lambda )$, is:
\begin{equation}
\Delta ( \Lambda ) = \Delta_0 \exp\left(- \frac{1}{2}\int_\Lambda^{\Lambda_0} \frac{J (
    \omega )}{\omega^2} d \omega\right) \label{delta_l}
\end{equation}
with the spectral function given in Eq. (\ref{spectralFunctionSB}). A renormalized low energy term, $\Dren$, can be defined by
\begin{equation}
\Dren = \Delta_0 e^{- \int_{\Dren}^{\Lambda_0} \frac{J (
    \omega )}{\omega^2} d \omega}\;.
\label{Dren}
\end{equation}
The free energy is again determined by Eq. (\ref{int_ener}), though cannot be
evaluated analytically, anymore. The scaling behavior of the renormalized tunneling given in Eq. (\ref{delta_l}) is no longer a power
law, as in the Ohmic case. Still, we can distinguish two limits:

i) The renormalization of $\Delta ( \Lambda )$ is slow. In this case,  the integral
in Eq. (\ref{int_ener}) is dominated by the region $\Lambda \sim \Lambda_0$,
where the function in the integrand goes as $\Lambda^{-2}$. The integral is
dominated by its higher cutoff, $\Lambda_0$, and the contribution from the
region near the lower cutoff, $\Dren$, can be neglected. Then, we obtain that
 $F ( \Delta_0 ) \sim \Delta_0^2 / \Lambda_0$.

ii) The renormalization of  $\Delta ( \Lambda )$ is fast. In this case, the contribution to the integral 
in Eq. (\ref{int_ener}) from the region $\Lambda \approx \Lambda_0$ is
small. The value of the integral is dominated by the region near $\Lambda
\simeq \Dren$. As $\Dren$ is the only quantity with dimensions of energy
needed to describe the properties of the system in this range, we expect that $F
( \Delta_0 ) \approx \Dren$.

In the scaling limit, $\Delta_0 /
\Lambda_0 \ll 1$, the values of the two terms, $\Dren$
and $\Delta_0^2 / \Lambda_0$, become very different. In
addition, there are no other energy scales which can qualitatively modify the
properties of the system. We thus conclude that only the two terms mentioned above will contribute to the free energy. Hence, we can write:
\begin{equation}
F ( \Delta_0 ) \simeq {\rm max} \left( \Dren , \frac{\Delta_0^2}{\Lambda_0}
\right)
\label{freeEnergy}
\end{equation}

In the following, we will use this conjecture to discuss super- and sub-Ohmic dissipation.
\begin{itemize}
\item[a)] {\it Super-Ohmic dissipation.} In the super-Ohmic case ($s > 1$), Eq. (\ref{Dren}) always has a solution and,
moreover, we can also set the lower limit of the integral to zero. This yields
\begin{equation}
\Dren = \Delta_0 e^{- \int_0^{\Lambda_0} \frac{J (
    \omega )}{\omega^2} d \omega} \approx \Delta_0 e^{- \alpha / ( s- 1 )}. 
\end{equation}
For $\alpha \gg 1$ we have $\Dren \ll \Delta_0$, but there is no transition from localized to delocalized behavior. 

Using Eq. (\ref{freeEnergy}) in the super-Ohmic case $s > 1$, we can approximately
write:
\begin{equation}
\langle \sigma_x \rangle \simeq {\rm max} \left( e^{- \alpha / ( s-1)} ,
  \frac{\Delta_0}{\Lambda_0} \right)
\end{equation}
We thus find a transition from underdamped to overdamped oscillations at some
critical coupling strength $\alpha \sim (s-1) \log ( \Lambda_0 / \Delta_0 )$.

It is finally interesting to note that the scaling analysis discussed
in Ref. \cite{K76} is equivalent to the scheme used here.

\item[b)] {\it Sub-Ohmic dissipation.} In the sub-Ohmic case ($s < 1$), it is not guaranteed that Eq. (\ref{Dren}) has a solution. In general, a solution only exists when $\Delta_0 / \Lambda_0$ is
not much smaller than 1. 

The existence of a phase transition in case of a sub-Ohmic bath was first proved in Ref. \cite{Spohn85}. Whereas the relation in Eq. (\ref{Dren}) and a similar analysis based on flow equations for Hamiltonians \cite{KM96} yields a discontinuous transition between the localized and delocalized
regimes, detailed numerical calculations suggest that the transition is continuous \cite{Bulla03}. 

Since there is a phase transition from localized to non-localized behavior,
there might also be a 
transition between overdamped to underdamped oscillation. 
In Ref. \cite{SM02}, this transition was discussed on the basis of spectral 
functions analogous to the discussion of Refs. \cite{G85,Costi96} for Ohmic
dissipation. 
It was found that for $s>0.5$ the transition takes place for lower 
values of $\alpha$ as in the Ohmic case, e.g., for $s=0.8$ and $\Lambda_0/\Delta_0=10$ the transition coupling strength is $\alpha^*\approx0.2$. For a recent discussion on the spectral properties using the Numerical Renormalization Group, see Ref. (\cite{Anders07}). 

Using Eqs. (\ref{Dren}) and (\ref{freeEnergy}) yields for the sub-Ohmic case the following qualitative behavior:
\begin{equation}
\langle \sigma_x \rangle \simeq \left\{ \begin{array}{llr} 1 &{\rm delocalized
      \, \, \,
      regime \, \, \,} &\frac{\Delta_0}{\Lambda_0} \simeq 1  \\
      \frac{\Delta_0}{\Lambda_0} &{\rm localized \, \, \, regime \, \, \,} 
      &\frac{\Delta_0}{\Lambda_0} \ll 1
      \end{array} \right.
\end{equation}
The analysis used in the previous cases leads us to expect coherent
oscillations in the delocalized regime.

We can extend the study of the sub-Ohmic case to the vicinity of the second
order phase transition described in Ref. \cite{VTB05}, which in our notation takes place for $\alpha=s\Delta_0 / \Lambda_0\ll 1$. In this regime, which
cannot be studied using  the Franck-Condon like renormalization of
Eq. (\ref{Dren}), we use the renormalization scheme around the fully coherent
state proposed in Ref. \cite{VTB05}. To one-loop order, the beta-function for the dimensionless
quantity (expressed in our notation) $\tilde{\kappa} = ( \alpha \Lambda ) / \Delta$ then reads 
\begin{equation}
\beta(\tilde{\kappa}) = -s\tilde{\kappa}+\tilde{\kappa}^2.
\label{beta_k}
\end{equation}
Near the transition, in the delocalized phase, $\tilde{\kappa}$ thus scales towards zero as
\begin{equation}
\tilde{\kappa} ( \Lambda ) = \tilde{\kappa}_0 \left(
  \frac{\Lambda}{\Lambda_0} \right)^s\;.
\label{scaling_k}
\end{equation}
The scaling of $\langle \sigma_x \rangle$ is
\begin{equation}
\frac{\partial \langle \sigma_x \rangle}{\partial \Lambda} = - \tilde{\kappa}
( \Lambda ) \frac{\Delta}{\Lambda^2}\;.
\label{scaling_sx}
\end{equation}
The fact that the scheme assumes a fully coherent state as a starting point
implies that $\Delta$ is not renormalized. Inserting Eq. (\ref{scaling_k})
into Eq. (\ref{scaling_sx}), we find:
\begin{equation}
\frac{\partial \langle \sigma_x \rangle}{\partial \Lambda} = -  \tilde{\kappa}_0 \left(
  \frac{\Lambda}{\Lambda_0} \right)^s \frac{\Delta_0}{\Lambda^2}
\end{equation}
\begin{figure}[t]
  \begin{center}
    \includegraphics*[width=3.5in,angle=0]{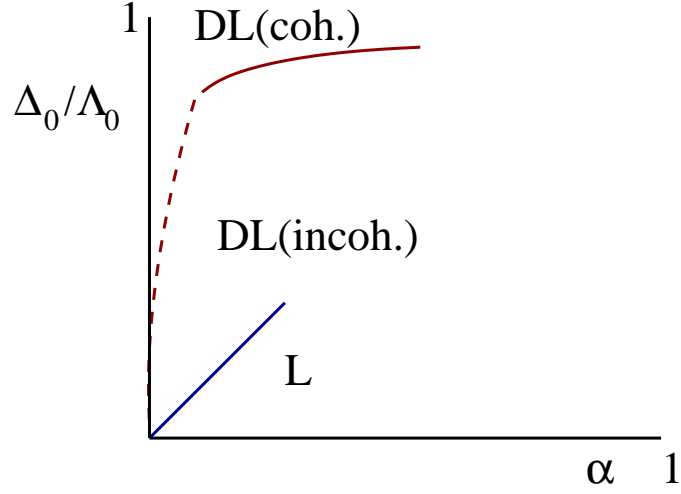}
    \caption{Schematic picture of the different regimes in the sub-Ohmic
      dissipative TLS studied in the text. DL stands for the delocalized phase,
      while L denotes the localized phase. The lower blue line denotes the
      continuous transition studied in Ref. \protect{\cite{VTB05}}. The red line
      marks the boundaries of a regime characterized by a small
      renormalization of the tunneling rate, Eq. (\protect{\ref{Dren}}), and
      coherent oscillations.}
    \label{subohmic}
\end{center}
\end{figure}

If we calculate $\langle \sigma_x \rangle$ from this equation, we find
that the resulting integral diverges as $\Lambda \rightarrow 0$ for $s \leq
1$. This result implies that $\langle \sigma_x \rangle \ll 1$. For
sufficiently low values of the effective cutoff, $\Lambda$, the value of
$\langle \sigma_x \rangle$ can be calculated using a perturbation expansion
on $\Delta_0$, leading to $\langle \sigma_x \rangle \sim \Delta_0 /
\Lambda_0$. This result implies the absence of coherent oscillations. A schematic picture of the regimes studied for the sub-Ohmic TLS is shown in
Fig. [\ref{subohmic}]. 
\end{itemize}

We finally note that the entanglement of a spin-1/2 particle coupled to a sub-Ohmic environment has recently been discussed in Ref. \cite{Hofstetter07}.
\section{Non-local information measure}
Quantum measurement is closely connected with the collapse of the wave function and due to the recent advances in quantum engineering, the concept of ``information'' has to be reconsidered when one deals with quantum mechanical systems. But instead of introducing a new concept of quantum information ``from scratch'', one can also start with the measuring process and see what information can be extracted. This line was recently pursued by Zurek and coworkers \cite{OPZ05} by proposing that in a classical description, information can be obtained by measuring the environment to which it is coupled. This approach seems even more appropriate for quantum mechanical systems.

\subsection{The model and information measure}

\begin{figure}[t]
  \begin{center}
    \includegraphics*[width=3in,angle=0]{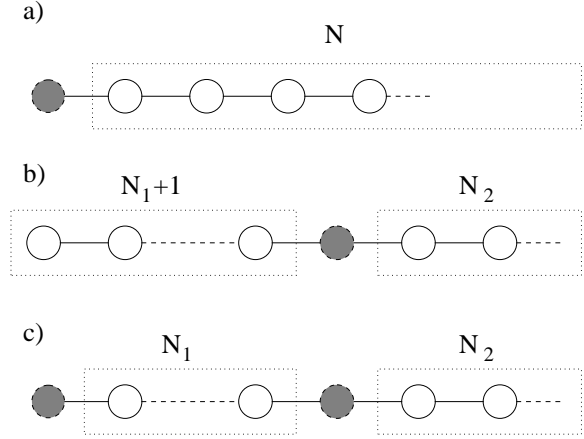}
    \caption{a) Measuring the system at the left end of the environment. b) Measuring part of the environment. c) Measuring the system and part of the environment simultaneously.}
    \label{Rubinmodel}
\end{center}
\end{figure}

In this section, we want to employ an information measure based on the
measurement process and apply it to a dissipative quantum system. The
model will consist of a harmonic chain with open boundaries. If
the mass of the first (quantum mechanical) particle is large compared
to the other masses, one speaks of the Rubin model \cite{R64}, but for
simplicity, we will choose all masses equally, here. In subsection
\ref{Ebetweentwoends}, we will then distinguish between the spring
constant of the bulk and of the edge. The Hamiltonian is given by
\begin{align}
\label{LinearChain}
    H=\sum_{n}\left(\frac{p_n^2}{2m}+\frac{f}{2}(x_{n+1}-x_{n})^2\right)\;.
\end{align}
The particle at the left end of the chain shall denote our system which is coupled to the environment $B_1$ (filled and empty circles in Fig. \ref{Rubinmodel}a), respectively).
\begin{itemize}
\item[a)] Measuring the system means that one is only interested in the mean value of the environment. The relevant density matrix is thus obtained by tracing out the bath degrees of freedom. The von Neumann entropy is known to be a good measure to characterize the ground state. We thus have
\begin{align}
E_1(\psi)=-\text{Tr}(\rho_1\ln\rho_1)\quad,\quad\rho_1=\text{Tr}_{B_1}(|\psi\rangle\langle\psi|)\quad.
\label{entropy1}
\end{align} 
The above model can be mapped to the dissipative harmonic oscillator with Ohmic coupling. This is done by diagonalizing the bath modes  
\begin{align}
x_n=\sqrt{\frac{2}{\pi}}\int_0^\pi dk\sin(kn)x(k)
\end{align}
and results in the following representation of the Hamiltonian:
\begin{align}
H=\frac{p^2}{2m}+\frac{f}{2}x+\sum_{k}\left(\frac{p_k^2}{2m}+\frac{m\omega_k^2}{2}x_k^2\right)+x\sum_k\lambda_kx_k
\label{DissipativeOscillator}
\end{align}
\item[b)] In order to apply the information approach proposed in Ref. \cite{OPZ05}, we will now pick out one of the environmental particles, see Fig. \ref{Rubinmodel}b). Again, the dissipative system (\ref{LinearChain}) can be brought into more familiar form by diagonalizing the left and the right part of the environment separately. This formally results in the problem where a quantum mechanical particle in a harmonic potential is coupled to {\em two} baths. But since the left and the right reservoir linearly couple to the same spatial coordinate, they are indistinguishable. The resulting model is thus the standard dissipative harmonic oscillator with modified coupling functions $\lambda_k$ as given by Eq. (\ref{DissipativeOscillator}).

The relevant density matrix for the selected particle of the environment is now obtained by tracing out the bath degrees of freedom without the selected particle plus the system itself, labeled as $B_2$. For the von Neumann entropy we thus have
\begin{align}
E_2(\psi)=-\text{Tr}(\rho_2\ln\rho_2)\quad,\quad\rho_2=\text{Tr}_{B_2}(|\psi\rangle\langle\psi|)\quad.
\label{entropy2}
\end{align}
\item[c)] The last step is to measure both, the system at the left end of the chain and the selected particle of the bath, see Fig. \ref{Rubinmodel}c). Again, we proceed by decoupling the left and right part of the environment separately. We obtain the following representation of the Hamiltonian in Eq. (\ref{LinearChain}):
\begin{align}
\notag
H&=\frac{p_1^2+p_2^2}{2m}+\frac{f}{2}(x_1^2+x_2^2)+\sum_{k,i=1,2}\left(\frac{p_{k,i}^2}{2m}+\frac{m\omega_{k,i}^2}{2}x_{k,i}^2\right)\\
&+x_1\sum_k\lambda_{k,1}x_{k,1}+x_2\sum_k\left(\kappa_{k,1}x_{k,1}+\kappa_{k,2}x_{k,2}\right)
\end{align}
In contrary to case b), here there is a distinction between the two resulting non-interacting reservoirs since one bath is coupled to two particles whereas the other bath only couples to the environmental particle. There is no way of preforming a unitary transformation such that the two reservoirs act as one. 

A similar type of problem has been analyzed by Kohler and Sols where two different baths were coupled to the momentum and to the spatial coordinate, respectively \cite{KS05}. Also from the two-channel Kondo model it is known that two baths can significantly alter the system behavior due to the simultaneous measurement process.\cite{Potok06} We thus expect that effects of quantum frustration are contained in the employed information measure.

The von Neumann entropy of the subsystem is given by tracing out the degrees of freedom of the remaining bath $B_3$
\begin{align}
E_3(\psi)=-\text{Tr}(\rho_3\ln\rho_3)\quad,\quad\rho_3=\text{Tr}_{B_3}(|\psi\rangle\langle\psi|)\quad.
\label{entropy3}
\end{align}
\end{itemize}

The measure of information which is contained by measuring parts of the environment as proposed by Ref. \cite{OPZ05} is now given by
\begin{align}
E(\psi)=E_3(\psi)-E_1(\psi)-E_2(\psi)\;.
\end{align}

In the following, we will set $N_2=0$, i.e., we investigate the entanglement between the two ends. In the context of spin-models, the long-distance entanglement was recently considered using as measure of entanglement the concurrence \cite{CamposVenuti06}. On the other hand, it was shown that the above measure based on the von Neumann entropy only captures classical correlations if it is positive \cite{Cerf97}.

\subsection{Entanglement between the two ends}
\label{Ebetweentwoends}
For explicit calculations, we will consider a simplified version of the above model and neglect the reservoir to the right, i.e., we will set $N_2=0$ in Fig. \ref{Rubinmodel} b), c). This amounts to the following question: What is the entanglement between the two ends of a harmonic chain as function of the system size $N$. 

The chain is confined by the masses at $x_0$ and $x_N$. The diagonalization of the harmonic chain for finite length $N-1$ yields
\begin{align}
x_n=\frac{1}{\sqrt{N-1}}\sum_{m=1}^{N-1}\sin(k_mn)x_{m}\quad,\quad k_m=\frac{\pi m}{N}\;.
\end{align}
The eigenvalues are given by $\omega_{m}^2=4f_B\sin^2(k_m/2)$. Here, we have introduced an extra spring constant $f_B$ for the masses of the bath to contrast it from the spring constant that connects the two masses at the end with the chain, denoted by $f$.

In the following, we will neglect finite size effects and only consider
the case where there are two particles at the end. The case of one
particle is then simply obtained by neglecting the second particle and the
transformed Hamiltonian reads 
\begin{align}
H&=\sum_{i=1,2}\frac{p_i^2}{2}+\frac{f}{2}q_i^2+\sum_m \Big(\frac{p_{m}^2}{2}+\frac{\omega_{m}^2}{2}x_{m}
-\sum_{i=1,2}c_m^ix_mq_i\Big)\;,
\end{align}
with ($N-1\rightarrow N$)
\begin{align}
c_m^1=\frac{f_B}{\sqrt{N}}\sin(k_m)\;,\;c_m^2=(-1)^m\frac{f_B}{\sqrt{N}}\sin(k_m)
\label{CouplingCoeff}
\end{align}
and $q_{i=1}=x_0$ and $q_{i=2}=x_N$.

To obtain the von Neumann entropy of the various subsystems listed in a)-c), we first need to compute the reduced density matrix. The reduced density matrix of dissipative systems is commonly represented as a path integral where the bath degrees of freedom have been integrated out \cite{Cal83,W99}:
\begin{align}
\langle q''|\rho_A|q'\rangle\propto\int_{q(0)=q'}^{q(\beta)=q''}{\mathcal D}q\exp(-S_S^{(E)}[q]-S_{infl}^{(E)}[q])
\end{align}
Here $S_S^{(E)}$ denotes the Euclidean action of the system and $S_{infl}^{(E)}$ the influence on the system due to the environment.
For one particle coupled on a linear chain with coupling constant $c_m$, we have
\begin{align}
S_{infl}^{(E)}[q]=-\sum_m\frac{c_m^2}{2}\frac{1}{\beta}\sum_n\frac{|q_n|^2}{\nu_n^2+\omega_m^2}
\end{align}
with the Fourier transform 
\begin{align}
q(\tau)=\frac{1}{\beta}\sum_nq_n\exp(i\nu_n\tau)\quad,\quad \nu_n=2\pi n/\beta\quad.
\end{align}
For two particles coupled to both ends of a linear chain with coupling constants $c_m^{(1/2)}$, we have
\begin{align}
S_{infl}^{(E)}[q]=-\sum_m\frac{1}{2}\frac{1}{\beta}\sum_n\frac{|c_m^1q_n^1+c_m^2q_n^2|^2}{\nu_n^2+\omega_m^2}\quad.
\end{align}
Notice that there is no potential renormalization in our model originating from the harmonic chain.

Since we have already discussed the von Neumann entropy for a
dissipative particle in a harmonic potential, we are left with the
case of two particle, see Fig. c).  With the coupling coefficient of
Eq. (\ref{CouplingCoeff}), the effective action $S=S_S^{(E)}+S_{infl}^{(E)}$ can be written as
$S=S_1+S_2+S_{int}$ with
\begin{align}
S_i&=\sum_{i=1,2}\frac{1}{2}\frac{1}{\beta}\sum_n (\nu_n^2+\omega_0^2-\omega_r^2(\nu_n))|q_n^i|^2
\end{align}
the effective action of particle $i$ coupled to the dissipative environment and
\begin{align}
S_{int}&=\frac{1}{\beta}\frac{1}{N}\sum_n\omega_I^2(\nu_n)\text{Re}(q_n^1q_n^2)
\end{align}
the effective action describing the interaction between the two particles through the environment. In the above equations, we further defined the potential renormalization $\omega_r(\nu_n)$ and the (system-size independent) effective splitting parameter $\omega_I(\nu_n)$ as 
\begin{align}
\omega_r^2(\nu_n)&=\frac{1}{N}\sum_{m=1}^{N-1}\frac{f_B^2\sin^2(k_m)}{\nu_n^2+4f_B\sin^2(k_m/2)}\quad,\\
\omega_I^2(\nu_n)&=-\sum_{m=1}^{N-1}(-1)^m\frac{f_B^2\sin^2(k_m)}{\nu_n^2+4f_B\sin^2(k_m/2)}\quad.
\end{align}

By a unitary transformation, $q_n^\pm=(q_n^1\pm q_n^2)/\sqrt{2}$, the two modes can be decoupled, i.e., $S=S_++S_-$: 
\begin{align}
S_\pm&=\sum_{i=1,2}\frac{1}{2}\frac{1}{\beta}\sum_n (\nu_n^2+\omega_0^2-\omega_r^2(\nu_n)\pm\omega_I^2(\nu_n)/N)|q_n^\pm|^2
\end{align}

The physical behavior of dissipative models is determined by the low-frequency modes of the bath. The action can thus be interpreted as the action of two harmonic oscillators with the effective frequencies $\omega_\pm=\sqrt{\tilde\omega^2\pm(\omega_I/N)^2}$ where $\tilde\omega=\sqrt{\omega_0^2-\omega_r^2}$. For $\nu_n\rightarrow0$, we further have $\omega_r^2=\omega_I^2=f_B/2$. 

For the chain with equal spring constant $f_B=f$, we have $\omega_0^2=f/2=f_B/2$ and thus $\tilde\omega=0$, which indicates a phase-transition to a localized state. For $f>f_B$, we can use the results of the entropy of an harmonic oscillator. In the expression of the entropy Eq. (\ref{EntropieHarmOsci}), only the combination $\langle q^2\rangle\langle p^2\rangle$ enters, such that the only dependence on the system size $N$ is contained in the term 
\begin{align}
\langle q^2\rangle\langle p^2\rangle\to\frac{\alpha f(\alpha)}{\pi}\ln \frac{\omega_c}{\omega_\pm}\quad.
\end{align}
Expanding the logarithm, the linear term cancels and we thus have for the information measure for two particles at the end of a harmonic chain with length $N$ the following scaling behavior:
\begin{align}
I&=E(A+B)-E(A)-E(B)\propto\frac{1}{N^2}
\end{align}

\section{Summary}
In this article, we have investigated the entanglement of quantum systems at the boundary. We have first calculated the entanglement between qubits at the
boundary of a spin chain, whose parameters are tuned to be near a
quantum critical point. The calculations show a behavior which
significantly differs from that inside the bulk of the chain.
Although the spins are part of the critical chain, we find no
signs of the scaling behavior which can be found in the bulk. We
use the same approach as done previously for 
bulk spins\cite{Oetal02,ON02}, although it should be noted that the
existence of a finite order parameter in the ordered phase will
change these results if the calculations are performed in the
presence of an infinitesimal applied field.

We have also considered the entanglement between two Ising-coupled
spins connected to a dissipative environment and which undergo a local
quantum phase transition. The system which we have studied belongs to
the generic class of systems with a Kosterlitz-Thouless transition at
zero temperature, like the Kondo model or the dissipative two level
system. The most remarkable feature of our results is that the
entanglement properties show a pronounced change at the parameter
values where the coherent quantum oscillations between the qubits are
lost.

In the second part of this article, the entanglement properties of
dissipative systems were investigated using the von Neumann
entropy. We first discuss two integrable dissipative quantum systems -
the free dissipative particle and the dissipative harmonic oscillator
- and calculated the von Neumann entropy. In the former case, we found
an analogy to the entropy of a canonical ensemble at temperature
$T$. The case of the harmonic oscillator is the more interesting one
since it exhibits a transition from underdamped to overdamped
oscillations. This transition is also manifested in the entropy, but
not as strongly as in the case of the spin-boson model. This is
probably due to the absence of a quantum phase transition and that the
model can be adequately treated by semi-classical methods as done
e.g. in the context of the fluctuation-dissipation theorem \cite{W99}.

We also calculated the von Neumann entropy for the spin-boson model on
the basis of a scaling approach for the free energy. Only in the Ohmic
case, the resulting integral, i.e., the ground-state energy, could be
evaluated and we analyzed the behavior at the transition from
underdamped to overdamped oscillations. We found that the change of
the logarithm of $\langle \sigma_x\rangle$ with respect to the
coupling strength $\alpha$ is strongly pronounced at the Toulouse
point. In the non-Ohmic case, we argued that the crossover between
coherent and decoherent oscillation takes place when the value of
$\langle \sigma_x \rangle$ becomes comparable to the result obtained
using a perturbation expansion in the tunneling matrix (as it is the
case for Ohmic dissipation). In this framework, we can also discuss the
super-Ohmic and sub-Ohmic dissipative two-level system,
respectively. We conclude that entanglement properties are closely
connected to the transition of coherent to incoherent tunneling.

In the third part of this paper, we have applied an extended measure
of quantum information to a simple model, describing a chain of
harmonically coupled particles. We argued that this measure can be
applied to relate particles of arbitrary distance (or arbitrary
regions of the chain) and that it incorporates features of quantum
frustration. We calculated explicitly the information measure which
relates the two particles at the two ends of the harmonic chain which
decays algebraically with the system size.

\section{Acknowledgments}
Funding from FCT (Portugal) grant PTDC/FIS/64404/2006 and from MEC
(Spain) grant FIS2004-06490-C03-01 is acknowledged.  \appendix
\section{Jordan-Wigner and Bogoljubov Transformation}
\label{App:Bogoljubov}
In this appendix, we start the discussion with the slightly more general
anisotropic spin-1/2 Heisenberg model in a homogeneous magnetic field, which is given by
\begin{align}
H=\sum_i\left\{J_xs_i^xs_{i+1}^x+J_ys_i^ys_{i+1}^y+J_zs_i^zs_{i+1}^z-hs_i^z\right\}\;,
\end{align}
where $s_i^\alpha\to\sigma_i^\alpha/2$, $\sigma^\alpha$ denoting the Pauli matrices with $\alpha=x,y,z$.

Introducing the new operators $a_i=s_i^x+is_i^y$ (which leads to $s_i^z=a_i^\dagger a_i-1/2$), one now performs a Jordan-Wigner transformation \cite{JW28}
\begin{align}
c_i=\exp\left\{i\pi\sum_{j=1}^{i-1}a_j^\dagger a_j\right\}a_i\quad,\quad 
a_i=\exp\left\{-i\pi\sum_{j=1}^{i-1}c_j^\dagger c_j\right\}c_i\quad.
\end{align}
With now anti-commuting $c_i$-operators $\{c_i,c_j^\dagger\}=\delta_{i,j}$, the Hamiltonian can thus be written in Fermion operators as
\begin{align}
H=\sum_i\left\{\frac{J_x+J_y}{4}(c_i^\dagger c_{i+1}+h.c.)\frac{J_x-J_y}{4}(c_i^\dagger c_{i+1}^\dagger+h.c.)+J_z(c_i^\dagger c_i-1/2)(c_{i+1}^\dagger c_{i+1}-1/2)-h(c_i^\dagger c_i-1/2)\right\}\;.
\end{align}
For the Ising model in a transverse field we set $J_y=J_z=0$ which yields
\begin{align}
H=\frac{J_x}{4}\sum_i\left[c_i^\dagger c_{i+1}+c_i^\dagger c_{i+1}^\dagger+h.c.\right]-h\sum_ic_i^\dagger c_i+Nh/2\;,
\end{align}
where we chose fixed boundary conditions since $a_N^\dagger a_1\neq
c_N^\dagger c_1$. For general boundary conditions, i.e., $\kappa\neq0$
in Eq. (\ref{IsingGen}), we will neglect the boundary term that involves the
operator $\exp(i\pi\sum_{i=1}^N c_i\dag c_i)$ in order to preserve the
bilinearity of the model, see Ref. \onlinecite{LSM61}. Using a more general notation for a bilinear Hamiltonian 
\begin{align}
H=\sum_{i,j=1}\left[A_{i,j}c_i^\dagger c_j+\frac{1}{2}\left(B_{i,j}c_i^\dagger c_j^\dagger+h.c\right)\right]\;,
\end{align}
the formal diagonalizing of the above Hamiltonian via a Bogoljubov transformation \cite{B47} leads to 
\begin{align}
H=\sum_{\alpha=1}^N\omega_\alpha\eta_\alpha^\dagger\eta_\alpha+\frac{1}{2}\left[\sum_{i=1}^N A_{i,i}-\sum_{\alpha=1}^N\omega_\alpha\right]\;,
\end{align}
where $\eta_\alpha=\sum_i g_{\alpha,i} c_i+h_{\alpha,i} c_i^\dagger$ are the operators in the diagonal basis.

The Bogoljubov transformation, i.e., the determination of the new eigenenergies $\omega_\alpha$ as well as the new operators through $g_{\alpha,i}$ and $h_{\alpha,i}$, is equivalent to solving the eigenvalue problem of the matrix $(A-B)(A+B)$. For the Ising model with open boundary condition, this is equivalent to the problem of a one-dimensional chain with an impurity at the first site, i.e., $(A-B)(A+B)\to(J_x/4)^2H_{TB}$ with 
\begin{align}
H_{TB}=-\sum_{i=1}^{N-1}2\mu (t_i^\dagger t_{i+1}+h.c.)+\mu^2t_1^\dagger t_1+(\mu^2+4)\sum_{i=2}^Nt_i^\dagger t_i
\end{align}
where $\mu=4h/J_x$.

We now want to analyze the eigenvectors and eigenenergies of the tight-binding model. For $1<n<N$, the eigenvectors $x_i$ and eigenenergies $\tilde\omega_\alpha^2=\omega_\alpha^2/(J_x/4)^2$ are given by the following equations:
\begin{align}
-2\mu x_{n-1}+(\mu^2+4-\tilde\omega_\alpha^2)x_n-2\mu x_{n+1}=0
\end{align}
With the Ansatz $x_n\sim e^{ik_\alpha n}$, we have for the extended states
\begin{align}
\tilde\omega_\alpha^2=\mu^2+4-4\mu \cos k_\alpha\;.
\end{align}
We are interested in the limit $N\to\infty$ where boundary conditions can be disregarded since $k$ will be continuous. For $\mu>0$, we then have $\tilde\omega_{max}=2+\mu$, $\tilde\omega_{min}=|2-\mu|$. 

For $\mu=2$, the continuum starts at zero energy which represents the critical point. For $\mu<2$, there is an additional ``bound'' state, i.e., $x_n\sim e^{-\kappa n}$. For $n>1$, we then have
\begin{align}
\tilde\omega_b^2=\mu^2+4-2\mu(e^\kappa+e^{-\kappa})\;,
\end{align} 
and for $n=1$, we have  $\tilde\omega_b^2=\mu^2-2\mu e^{-\kappa}$. This leads to the solution $e^{-\kappa}=\mu/2$ and thus (for $N\to\infty$)
\begin{align}
\tilde\omega_b^2=0\;.
\end{align}
The restriction $\mu\leq2$ follows from the condition $\kappa>0$, i.e., a normalizable eigenfunction. The emergence of the bound state can be interpreted as a loss of coherence. Since it is connected to the appearance of a zero energy mode which is inherent to a quantum phase transition, we believe that this view point can be generalized to other quantum phase transitions.

With $\frac{1}{2}\sum_i A_{ii}=-hN/2$, we have for the ground-state energy
\begin{align}
E_0=-\frac{1}{2}\sum_{\alpha}\sqrt{J_x^2/4+h^2-hJ_x\cos k_\alpha}\;.
\end{align}
With the Hellmann-Feynman theorem $\langle m_z\rangle=-\frac{1}{N}\partial_h E_0$, we have for $N\to\infty$
\begin{align}
\langle m_z\rangle=\frac{1}{2\pi}\int_0^\pi dk\frac{h-\frac{1}{2}J_x\cos k}{J_x^2/4+h^2-hJ_x\cos k}\;.
\end{align}
At the critical point $h=h_c=J/2$, this leads to the logarithmic divergence of $\partial_h\langle m_z\rangle$.

For finite temperatures, we have
\begin{align}
F&=-kT\left[-\sum_\alpha\frac{\omega_\alpha}{2}+\sum_\alpha\ln (1+e^{-\beta\omega_\alpha})\right]
\to-kTN\int_0^\pi dk\ln\left[2\cosh(\beta\omega(k)/2)\right]\;.
\end{align}
With $\langle m_z\rangle=-\frac{1}{N}\partial_h F$, we have
\begin{align}
\langle m_z\rangle=\frac{1}{2\pi}\int_0^\pi dk\frac{h-\frac{1}{2}J_x\cos k}{J_x^2/4+h^2-hJ_x\cos k}\tanh\left[\frac{\beta}{2}\sqrt{J_x^2/4+h^2-hJ_x\cos k}\right]\;.
\end{align}
The singularity of $\partial_h\langle m_z\rangle$ at $h=h_c=J/2$ is thus suppressed for $T>0$.

\section{Calculation of the free energy of the dissipative TLS}
\label{App:Scaling}
We calculate the free energy of the dissipative two level system following the
scaling approach discussed for the Kondo problem in Refs. \cite{AYH70,AY71}, and formulated in a more general way in Ref. \cite{C81}. For the general long-ranged Ising model, the scaling approach was first applied by Kosterlitz \cite{K76}.

\begin{figure}[t]
  \begin{center}
    \includegraphics*[width=3in,angle=0]{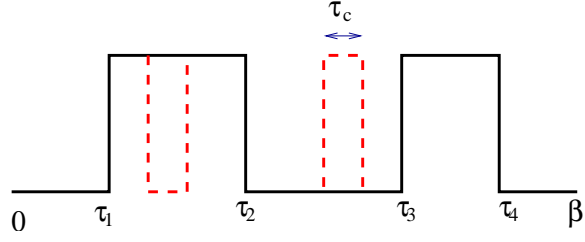}
    \caption{Sketch of the instanton pairs which renormalizes the calculation of the free energy of the dissipative TLS.}
    \label{free_energy}
\end{center}
\end{figure}

The partition function of the model can be expanded in powers of $\Delta^2$ as
\begin{equation}
Z = \sum_n \frac{\Delta^{2n}}{2n !} \int_0^\beta d \tau_1 
\cdots \int_0^\beta d
\tau_{2n} \prod_{ij=1,..,2n} f [ ( \tau_i - \tau_j ) / \tc ]
\label{free_ener}
\end{equation}
where $f [ ( \tau_i - \tau_j ) / \tc ]$ denotes the interaction between the kinks
located at positions $\tau_i$ and $\tau_j$. A term in the series is
schematically depicted in Fig. [\ref{free_energy}]. The scaling procedure
lowers the short time cutoff of the theory from $\tc$ to $\tc - d \tc$. This process
removes from each term in the sum in Eq. (\ref{free_ener}) details at times
shorter than $\tc - d \tc$. The rescaling $\tc \rightarrow \tc - d \tc$
implies the change $\Delta \rightarrow \Delta ( 1 + d \tc / \tc )$. The
dependence of $f [ ( \tau_i - \tau_j ) / \tc ]$ leads to another rescaling,
which can be included in a global renormalization of
$\Delta$ \cite{AYH70,AY71,C81}. In addition, configurations with an
instanton-antiinstanton pair at distances between $\tc$ and $\tc - d \tc$
have to be replaced by configurations where this pair is absent, as
schematically shown in Fig. [\ref{free_energy}]. The number of removed pairs
is proportional to $d \tc / \tc$. The center of the pair can be anywhere in
the interval $0 \le \tau \le \beta$. The final effect is the rescaling:
\begin{equation} 
Z \rightarrow Z \left( 1 + \Delta^2 \beta d \tc \right)
\label{ren_Z}
\end{equation}
Writing $Z$ as $Z = e^{- \beta F}$, where $F$ is the free energy,
Eq. (\ref{ren_Z}) can be written as:
\begin{equation}
- \frac{\partial F}{\partial \tc} = \Delta^2 ( \tc )
\end{equation}
In the Ohmic case, the dependence of $\Delta$ on $\tc = \Lambda^{-1}$ is
\begin{equation}
\Delta ( \Lambda ) = \Delta_0 \left( \frac{\Lambda}{\Lambda_0} \right)^\alpha
\end{equation}
and, finally, we find the following relation:
\begin{equation}
\frac{\partial F}{\partial \Lambda} =  \left[ \frac{\Delta ( \Lambda
    )}{\Lambda} \right]^2 = \left( \frac{\Delta_0}{\Lambda_0} \right)^2
    \left( \frac{\Lambda}{\Lambda_0} \right)^{2 \alpha - 2}
\end{equation}
This equation ceases to be valid for $\Lambda \simeq \Delta_{\rm
  ren}$. For finite temperatures, we obtain
\begin{equation}
F ( T ) = \int_T^{\Lambda_0} \frac{\partial F}{\partial \Lambda} d \Lambda\;.
\label{int_free}
\end{equation}

It is interesting to apply this analysis to a free two level system. The
value of $\Delta_0$ does not change under scaling. We find the following expression:
\begin{equation}
\frac{\partial F}{\partial \Lambda} = \left\{ \begin{array}{lr} \left( \frac{\Delta_0}{\Lambda}
\right)^2 &\Delta_0 \ll \Lambda \\ 0 &\Lambda \ll \Delta_0 \end{array} \right.
\end{equation}
Inserting this expression into Eq. (\ref{int_free}), we obtain
\begin{equation}
F ( T ) = \left\{ \begin{array}{lr} \frac{\Delta_0^2}{T} &\Delta_0 \ll T \\
    \Delta_0 &T \ll \Delta_0 \end{array} \right.
\end{equation}
and, finally:
\begin{equation}
\langle \sigma_x \rangle = \frac{\partial F}{\partial \Delta_0} = 
\left\{ \begin{array}{lr} \frac{\Delta_0}{T}  &\Delta_0 \ll T 
\\ 1 &T \ll \Delta_0 \end{array} \right.
\end{equation}
in qualitative agreement with the exact result $\langle \sigma_x \rangle =
\tanh ( \Delta_0 / T )$.
\bibliography{References}

\begin{thebibliography}{53}
\bibitem[{not()}]{note2}
\bibinfo{note}{It is interesting to note that the effective value of $\alpha$
  for the Ising model considered in the previous section is $\alpha = 1 / 2$,
  as, at criticality, $\langle \sigma_x ( t ) \sigma_x ( t' ) \rangle \sim ( t
  - t' )^{-1}$.}
\end{thebibliography}
\end{document}